\documentclass[runningheads]{llncs}
\usepackage{graphicx}
\usepackage{booktabs}  
\usepackage{amssymb}
\usepackage{amsmath}
\usepackage{hyperref}
\usepackage{multirow}

\newsavebox\CBox

\hypersetup{
    colorlinks=true,
    linkcolor=blue,
    filecolor=magenta,      
    urlcolor=blue,
}
\begin{document}
\title{Document Collection Visual Question Answering}
%
%
\author{Rub{\`e}n Tito \and Dimosthenis Karatzas \and Ernest Valveny}

%
\authorrunning{R. Tito et al.}
%
\institute{Computer Vision Center, UAB, Spain \\
\email{\{rperez, dimos, ernest\}@cvc.uab.es}}

\maketitle              
\begin{abstract}


Current tasks and methods in Document Understanding aims to process documents as single elements. However, documents are usually organized in collections (historical records, purchase invoices), that provide context useful for their interpretation.
To address this problem, we introduce Document Collection Visual Question Answering (DocCVQA) a new dataset and related task, where questions are posed over a whole collection of document images and the goal is not only to provide the answer to the given question, but also to retrieve the set of documents that contain the information needed to infer the answer. Along with the dataset we propose a new evaluation metric and baselines which provide further insights to the new dataset and task. 


\end{abstract}
\section{Introduction}

Documents are essential for humans since they have been used to store knowledge and information over the history. For this reason there has been a strong research effort on improving the machine understanding of documents. The research field of Document Analysis and Recognition (DAR) aims at the automatic extraction of information presented on paper, initially addressed to human comprehension. Some of the most widely known applications of DAR involve processing office documents by recognizing text \cite{hull1994database}, tables and forms layout \cite{couasnon2014recognition}, mathematical expressions \cite{mouchere2016icfhr2016} and visual information like figures and graphics \cite{siegel2016figureseer}. However, even though all these research fields have progressed immensely during the last decades, they have been agnostic to the end purpose they can be used for. Moreover, despite the fact that document collections are as ancient as documents themselves, the research in this scope has been limited to document retrieval by lexical content in word spotting \cite{manmatha1997word,rath2003word}, blind to the semantics and ignoring the task of extracting higher level information from those collections.


On the other hand, over the past few years Visual Question Answering (VQA) has been one of the major relevant tasks as a link between vision and language. Even though the works of \cite{biten2019scene} and \cite{singh2019towards} start considering text in VQA by requiring the methods to read the text in the images to answer the questions, they constrained the problem to natural scenes. It was \cite{mathew2020document} who first introduced VQA on documents. However, none of those previous works consider the image collection perspective, neither from real scenes nor documents.



\begin{figure}
\centering
\begin{tabular}{p{0.49\textwidth} p{0.49\textwidth}}
\includegraphics[width=\linewidth, keepaspectratio]{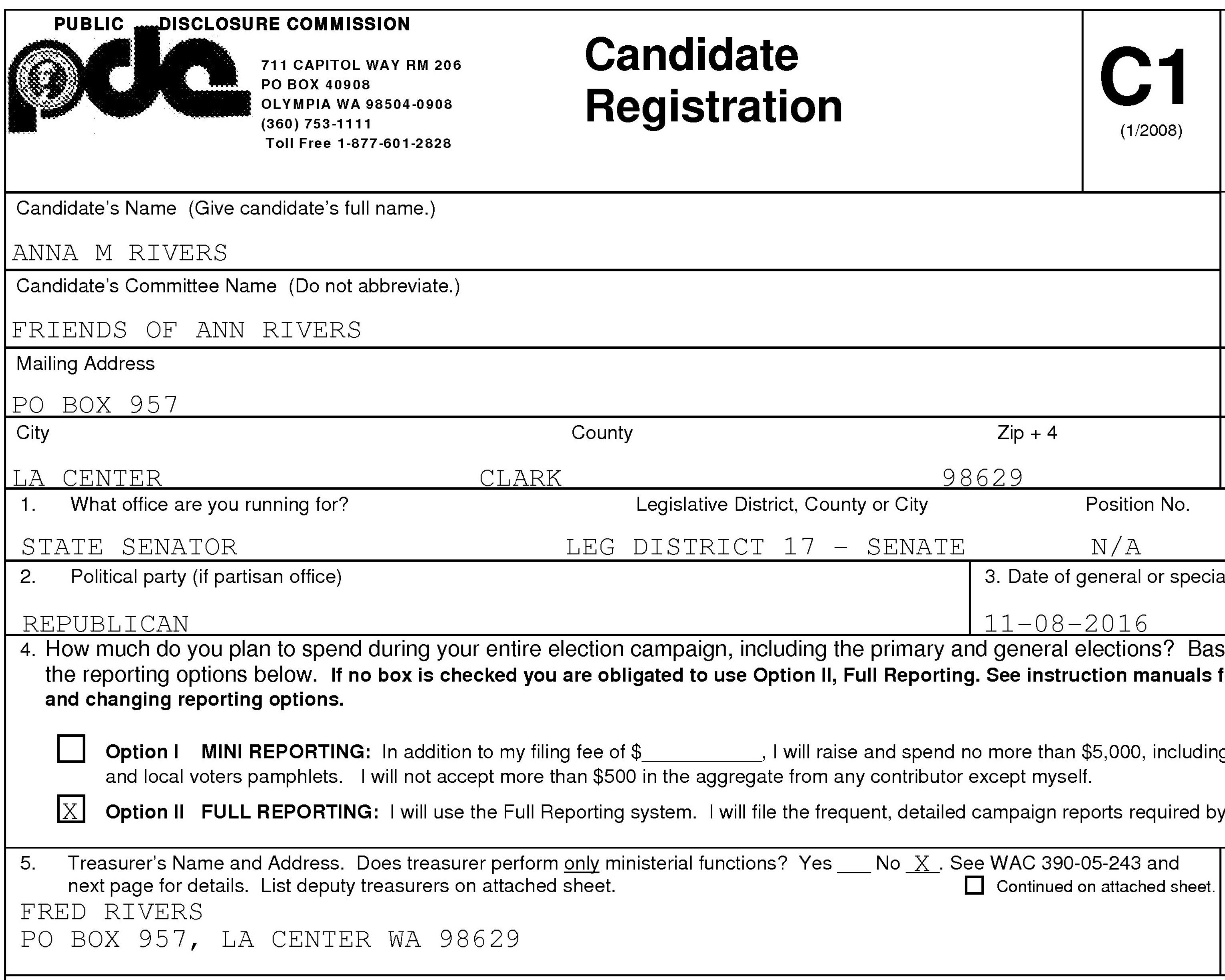} &
\includegraphics[width=\linewidth, keepaspectratio]{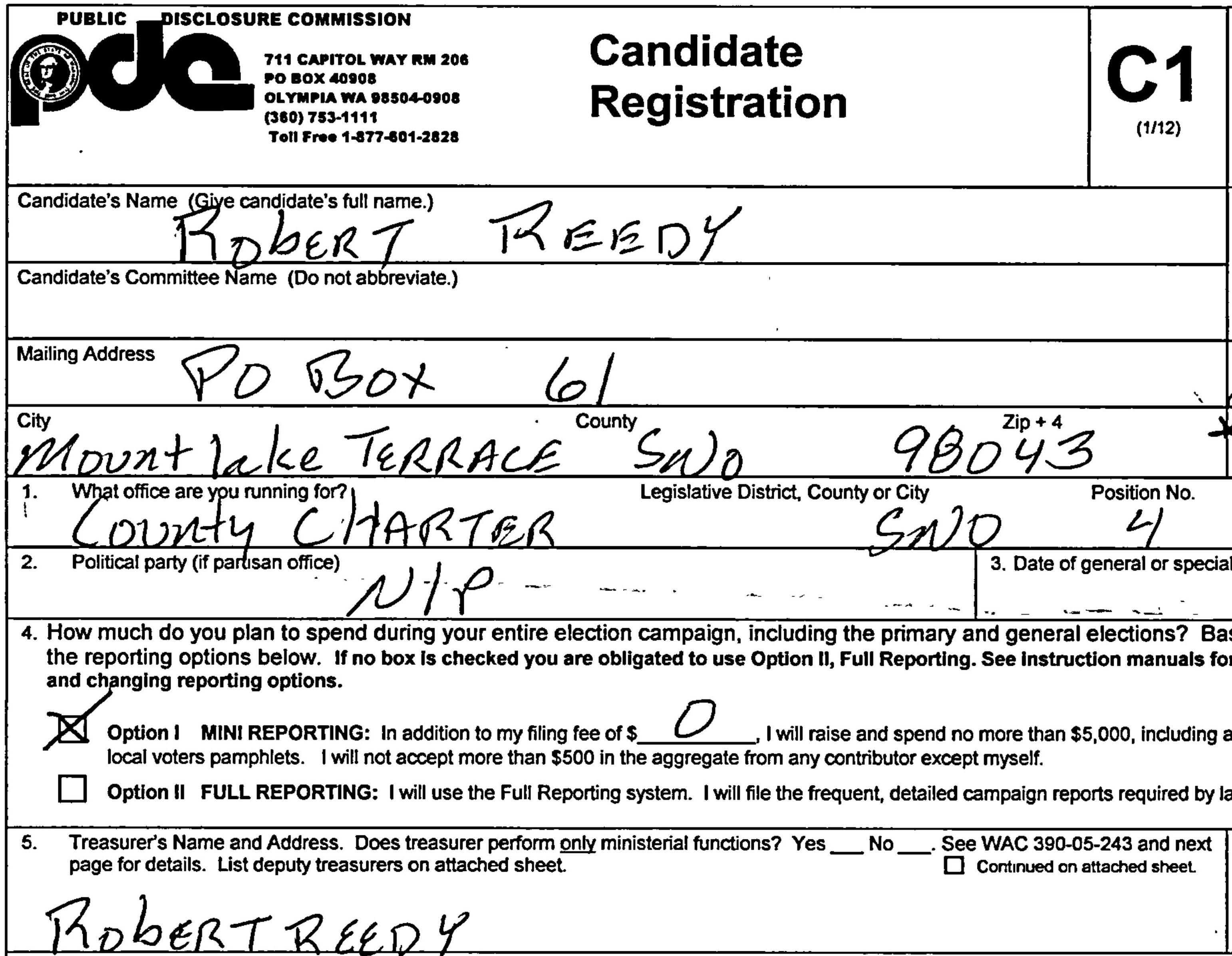} \\
\multicolumn{2}{l}{Q: In which years did Anna M. Rivers run for the State senator office?} \\
\multicolumn{2}{l}{A: [2016, 2020]} \\
\multicolumn{2}{l}{E: [454, 10901]} \\

\end{tabular}
\caption{Top: Partial visualization of sample documents in DocCVQA. The left document corresponds to the document with ID 454, which is one of the relevant documents to answer the question below. Bottom: Example question from the sample set, its answer and their evidences. In DocCVQA the evidences are the documents where the answer can be inferred from. In this example, the correct answer are the years 2016 and 2020, and the evidences are the document images with ids 454 and 10901 which corresponds to the forms where Anna M. Rivers presented as a candidate for the State senator office.}
\label{fig:example_docs}
\end{figure}

In this regard, we present Document Collection Visual Question Answering (DocCVQA) as a step towards better understanding document collections and going beyond word spotting. The objective of DocCVQA is to extract information from a document image collection by asking questions and expecting the methods to provide the answers. Nevertheless, to ensure that those answers have been inferred using the documents that contain the necessary information, the methods must also provide the IDs of the documents used to obtain the answer in the form of a confidence list as answer evidence. 
Hence, we design this task as a retrieval-answering task, for which the methods should be trained initially on other datasets and consequently, we pose only a set of 20 questions over this document collection. In addition, most of the answers in this task are actually a set of words extracted from different documents for which the order is not relevant, as we can observe in the question example in Figure \ref{fig:example_docs}. Therefore, we define a new evaluation metric based on the Average Normalized Levenshtein Similarity (ANLS) to evaluate the answering performance of the methods in this task. Finally, we propose two baseline methods from very different perspectives which provide some insights on this task and dataset. 

The dataset, the baselines code and the performance evaluation scripts with an online evaluation service are available in \href{https://docvqa.org}{https://docvqa.org}.

\section{Related Work}

\subsection{Document understanding}

Document understanding has been largely investigated within the document analysis community with the final goal of automatically extracting relevant information from documents. Most works have focused on structured or semi-structured documents such as forms, invoices, receipts, passports or ID cards, e-mails, contracts, etc. Earlier works~\cite{dengel2002,Schuster2013} were based on a predefined set of rules that required the definition of specific templates for each new type of document. Later on, learning-based methods~\cite{couasnon2014recognition,Palm2017} allowed to automatically classify the type of document and identify relevant fields of information without predefined templates. Recent advances on deep learning~\cite{liu2019,Xu2020,Zhang2020} leverage natural language processing, visual feature extraction and graph-based representations in order to have a more global view of the document that take into account word semantics and visual layout in the process of information extraction. 

All these methods mainly focus on extracting key-value pairs, following a bottom-up approach, from the document features to the relevant semantic information. The task proposed in this work takes a different top-down approach, using the visual question answering paradigm, where the goal drives the search of information in the document. 

\subsection{Document retrieval}

Providing tools for searching relevant information in large collections of documents has been the focus of document retrieval. Most works have addressed this task from the perspective of word spotting~\cite{rath2003word}, i.e., searching for specific query words in the document collection without relying on explicit noisy OCR. Current state-of-the-art on word spotting is based on similarity search in an common embedding space~\cite{almazan2014word} where both the query string and word images can be projected using deep networks~\cite{Krishnan2018,Sudholt2017} In order to search for the whole collection, these representations are combined with standard deep learning architectures for object detection in order to find all instances of a given word in the document~\cite{Krishnan2018,Wilkinson2017}. 

Word spotting only allows to search for the specific instances of a given word in the collection without taking into account the semantic context where that word appears. On the contrary, the task proposed in this work does not aim to find specific isolated words, but to make a semantic retrieval of documents based on the query question.

\subsection{Visual Question Answering}

Visual Question Answering (VQA) is the task where given an image and a natural language question about that image, the objective is to provide an accurate natural language answer. It was initially introduced in \cite{malinowski2014multi,ren2015exploring} and \cite{antol2015vqa} proposed the first large scale dataset for this task. All the images from those works are real scenes and the questions mainly refer to objects present in the images. Nonetheless, the field became very popular and several new datasets were released exploring new challenges like ST-VQA~\cite{biten2019scene} and TextVQA~\cite{singh2019towards}, which were the first datasets that considered the text in the scene. In the former dataset, the answers are always contained within the text found in the image while the latter requires to read the text, but the answer might not be a direct transcription of the recognized text. The incorporation of text in VQA posed two main challenges. First, the number of classes as possible answers grew exponentially and second, the methods had to deal with a lot of out of vocabulary (OOV) words both as answers or as input recognized text. To address the problem of OOV words, embeddings such as Fasttext~\cite{bojanowski2017enriching} and PHOC~\cite{almazan2014word} became more popular, while in order to predict an answer, along with the standard fixed vocabulary with the most common answers a copy mechanism was introduced by \cite{singh2019towards} which allowed to propose an OCR token as an answer. Later \cite{hu2020iterative} changed the classification output to a decoder that outputs a word from the fixed vocabulary or from the recognized text at each timestep, and provided more flexibility in complex and longer answers.

Concerning documents, FigureQA~\cite{kahou2017figureqa} and DVQA~\cite{kafle2018dvqa} focused on complex figures and data representation like different kinds of charts and plots by proposing synthetic datasets and corresponding questions and answers over those figures. More recently,~\cite{mathew2021docvqa} proposed DocVQA, the first VQA dataset over document images, where the questions also refer to figures, forms or tables but also text in complex layouts. Along with the dataset they proposed some baselines based on NLP and scene text VQA models. In this sense, we go a step further extending this work for document collections.

Finally, one of the most relevant works for this paper is ISVQA~\cite{bansal2020visual} where the questions are asked over a small set of images which consist of different perspectives of the same scene. Notice that even though the set up might seem similar, the methods to tackle this dataset and the one we propose are very different. For ISVQA all the images share the same context, which implies that finding some information in one of the images can be useful for the other images in the set. In addition, the image sets are always small sets of $6$ images, in contrast to the whole collection of DocCVQA and finally, the images are about real scenes which don't even consider the text. As an example, the baselines they propose are based on the HME-VideoQA~\cite{fan2019heterogeneous} and standard VQA methods stitching all the images, or the images features. Which are not suitable to our problem.


\section{DocCVQA Dataset} 

In this section we describe the process for collecting images, questions and answers, an analysis of the collected data and finally, we describe the metric used for the evaluation of this task. 

\subsection{Data Collection}



\subsubsection{Images:}
The DocCVQA dataset comprises $14,362$ document images sourced from the Open Data portal of the Public Disclosure Commission (PDC), an agency that aims to provide public access to information about the financing of political campaigns, lobbyist expenditures, and the financial affairs of public officials and candidates. We got the documents from this source for various reasons. First, it's a live repository that is updated periodically and therefore, the dataset can be increased in size in the future if it's considered necessary or beneficial for the research. In addition, it contains a type of documents in terms of layout and content that makes sense and can be interesting to reason about an entire collection. Moreover, along with the documents, they provide their transcriptions in the form of CSV files which allows us to pose a set of questions and get their answers without the costly process of annotation. From the original collection of document images, we discarded all the multi-page documents and documents for which the transcriptions were partially missing or ambiguous. Thus, all documents that were finally included in the dataset were sourced from the same document template, the US Candidate Registration form, with slight design differences due to changes over the time. However, these documents still pose some challenges since the proposed methods will need to understand its complex layout, as well as handwritten and typewritten text at the same time. We provide some document examples in Figure~\ref{fig:example_docs}. 



\subsubsection{Questions and Answers:} \label{QuestionsAndAnswers}

Considering that DocCVQA dataset is set up as a retrieval-answering task and documents are relatively similar we pose only a set of $20$ natural language questions over this collection. To gather the questions and answers, we first analyzed which are the most important fields in the document form in terms of complexity (numbers, dates, candidate's names, checkboxes and different form field layouts) and variability (see section \ref{DatasetStatistics}). We also defined different types of constraints for the types of questions since limiting the questions to find a specific value would place this in a standard word spotting scenario. Thus, we defined different constraints depending on the type of field related to the question: for dates, the questions will refer to the document before, after and between specific dates, or specific years. For other textual fields the questions refer to documents with specific values (candidates from party $P$), to documents that do not contain specific values (candidates which do not represent the party $P$), or that contains a value from a set of possibilities (candidates from parties $P$, $Q$ or $R$). For checkboxes we defined constraints regarding if a value is checked or not. Finally, according to the fields and constraints defined we posed the questions in natural language referring to the whole collection, and asking for particular values instead of the document itself. We provide the full list of questions in the test set in table~\ref{tab:questions}.

\setlength{\tabcolsep}{8pt}  
\renewcommand{\arraystretch}{1.5} 

\begin{table}[h]
\centering
\begin{tabular}{cp{10.5cm}}
ID  & Question                                                                                                         \\
8   & Which candidates in 2008 were from the Republican party?                                                         \\
9   & Which candidates ran for the State Representative office between 06/01/2012 and 12/31/2012?                      \\
10  & In which legislative counties did Gary L. Schoessler run for County Commissioner?                                \\
11  & For which candidates was Danielle Westbrook the treasurer?                                                       \\
12  & Which candidates ran for election in North Bonneville who were from neither the Republican nor Democrat parties? \\
13  & Did Valerie I. Quill select the full reporting option when she ran for the 11/03/2015 elections?                 \\
14  & Which candidates from the Libertarian, Independent, or Green parties ran for election in Seattle?                \\
15  & Did Suzanne G. Skaar ever run for City Council member?                                                           \\
16  & In which election year did Stanley J Rumbaugh run for Superior Court Judge?                                      \\
17  & In which years did Dean A. Takko run for the State Representative office?                                        \\
18  & Which candidates running after 06/15/2017 were from the Libertarian party?                                       \\
19  & Which reporting option did Douglas J. Fair select when he ran for district court judge in Edmonds? Mini or full? \\
    &
\end{tabular}
\caption{Questions in test set.}
\label{tab:questions}
\end{table}

\renewcommand{\arraystretch}{1} 
\setlength{\tabcolsep}{6pt}     

\begin{figure}
    \centering
    \includegraphics[width=\textwidth]{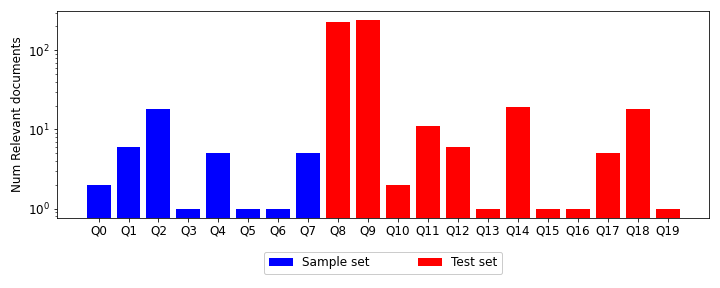}
    \caption{Number of relevant documents in ground truth for each question in the sample set (blue) and the test set (red).}
    \label{fig:rel_documents}
\end{figure}

Once we had the question, we got the answer from the annotations downloaded along with the images. Then, we manually checked that those answers were correct and unambiguous, since some original annotations were wrong. Finally, we divided the questions into two different splits; the sample set with $8$ questions and the test set with the remaining $12$. Given the low variability of the documents layout, we ensured that in the test set there were questions which refer to document form fields or that had some constraints that were not seen in the sample set. In addition, as depicted in figure~\ref{fig:rel_documents} the number of relevant documents is quite variable among the questions, which poses another challenge that methods will have to deal with.




\subsection{Statistics and Analysis} \label{DatasetStatistics}
We provide in table~\ref{tab:form_fields_desc} a brief description of the document forms fields used to perform the questions and expected answers with a brief analysis of their variability showing the number of values and unique values in their annotations.

\subsection{Evaluation metrics}

The ultimate goal of this task is the extraction of information from a collection of documents. However, as previously demonstrated, and especially in unbalanced datasets, models can learn that specific answers are more common to specific questions. One of the clearest cases is the answer \emph{Yes}, to questions that are answered with \emph{Yes} or \emph{No}. To prevent this, we not only evaluate the answer to the question, but also if the answer has been reasoned from the document that contains the information to answer the question, which we consider as evidence. Therefore, we have two different evaluations, one for the evidence which is based on retrieval performance, and the other for the answer, based on text VQA performance.

\setlength{\tabcolsep}{8pt}  

\begin{table}[h]
\centering
\begin{tabular}{llrr}
\textbf{Field}   & \textbf{Type} & \textbf{\# Values} & \textbf{\# Unique values} \\
Candidate name   & Text          & 14362              & 9309                      \\
Party            & Text          & 14161              & 10                        \\
Office           & Text          & 14362              & 43                        \\
Candidate city   & Text          & 14361              & 476                       \\
Candidate county & Text          & 14343              & 39                        \\
Election date    & Date          & 14362              & 27                        \\
Reporting option & Checkbox      & 14357              & 2                         \\
Treasurer name   & Text          & 14362              & 10197                     \\
                 &               &                    & 
\end{tabular}
\caption{Description of the document forms fields with a brief analysis of their variability showing the number of values and unique values in their annotations.}
\label{tab:form_fields_desc}
\end{table}

\setlength{\tabcolsep}{6pt}  

\subsubsection{Evidences:} Following standard retrieval tasks~\cite{liu2009learning} we use the Mean Average Precision (MAP) to assess the correctness of the positive evidences provided by the methods. We consider as positive evidences the documents in which the answer to the question can be found.

\subsubsection{Answers:} Following other text based VQA tasks~\cite{biten2019icdar,biten2019scene} we use the Average Normalized Levenshtein Similarity (ANLS) which captures the model's reasoning capability while smoothly penalizing OCR recognition errors. However, in our case the answers are a set of items for which the order is not relevant, in contrast to common VQA tasks where the answer is a string. Thus, we need to adapt this metric to make it suitable to our problem. We name this adaptation as Average Normalized Levenshtein Similarity for Lists (ANLSL), formally described in equation \ref{eq:anlsl}. Given a question $Q$, the ground truth list of answers $G = \{g_{1}, g_{2} \dots g_{M}\}$ and a model's list predicted answers $P = \{p_{1}, p_{2} \dots p_{N}\}$, the ANLSL performs the Hungarian matching algorithm to obtain a $k$ number of pairs $U = \{u_{1}, u_{2} \dots u_{K}\}$ where $K$ is the minimum between the ground truth and the predicted answer lists lengths. The Hungarian matching ($\Psi$) is performed according to the Normalized Levenshtein Similarity ($NLS$) between each ground truth element $g_{j} \in G$ and each prediction $p_{i} \in P$. Once the matching is performed, all the NLS scores of the $u_{z} \in U$ pairs are summed and divided for the maximum length of both ground truth and predicted answer lists. Therefore, if there are more or less ground truth answers than the ones predicted, the method is penalized.




\begin{equation} \label{eq:anlsl}
\begin{aligned}
U = {\Psi}(NLS(G, P)) \\
ANLSL = \frac{1}{\max(M,N)} \sum_{z=1}^{K}NLS(u_{z})  \\
\end{aligned}
\end{equation}

\section{Baselines}

This section describes the two baselines that are employed in the experiments. Both baselines breakdown the task into two different stages. First, they rank the documents according to the confidence of containing the information to answer a given a question and then, they get the answers from the documents with the highest confidence. The first baseline combines methods from the word spotting and NLP Question Answering fields to retrieve the relevant documents and answer the questions. We name this baseline as \emph{Text spotting + QA}. In contrast, the second baseline is an ad-hoc method specially designed for this task and data, which consist on extracting the information from the documents and map it in the format of key-value relations. In this sense it represents the collection similar as databases do, for which we name this baseline as \emph{Database}. These baselines allows to appreciate the performance of two very different approaches.


\subsection{Text spotting + QA}

The objective of this baseline is to set a starting performance result from the combination of two simple but generic methods that will allow to assess the improvement of future proposed methods.

\subsubsection{Evidence retrieval:} To retrieve the relevant documents we apply a text spotting approach, which consist on ranking the documents according to a confidence given a query, which in our case is the question. To obtain this confidence, we first run a Part Of Speech (POS) tagger over the question to identify the most relevant words in it by keeping only nouns and digits, and ignore the rest of the words. Then, as described in equation \ref{eq:confidence}, given a question $Q$, for each relevant word in the question $qw_{i} \in Q$ we get the minimum Normalized Levenshtein Distance ($NLD$)~\cite{levenshtein1966binary}, between all recognized words $rw_{j}$ extracted through an OCR 
in the document and the question word. Then, we average over all the distances and use the result as the confidence $c$ for which the document $d$ is relevant to the question.

\begin{equation} \label{eq:confidence}
c_{d} = \frac{1}{|Q|} \sum_{i=1}^{|Q|} \min_{j=1}^{|OCR|}\{NLD(qw_{i}, rw_{j})\}
\end{equation}

Notice that removing only stopwords is not enough, like in the question depicted in figure \ref{fig:example_docs}, where the verb \emph{run} is not considered as stopword, but can't be found in the document and consequently would be counterproductive.


\subsubsection{Answering:} Once the documents are ranked, to answer the given questions we make use of BERT~\cite{devlin2018bert} question answering model. BERT is a task agnostic language representation based on transformers~\cite{vaswani2017attention} that can be afterwards used in other downstream tasks. In our case, we use extractive question answering BERT models which consist on predicting the answer as a text span from a context, usually a passage or paragraph by predicting the start and end indices on that context. 
Nonetheless, there is no such context in the DocCVQA documents that encompasses all the textual information. Therefore, we follow the approach of \cite{mathew2021docvqa} to build this context by serializing the recognized OCR tokens on the document images to a single string separated by spaces following a top-left to bottom-right order. Then, following the original implementation of \cite{devlin2018bert} we introduce a start vector $S \in \mathbb{R}^{H}$ and end vector $E \in \mathbb{R}^{H}$. The probability of a word $i$ being the start of the answer span is obtained as the dot product between the BERT word embedding hidden vector $T_{i}$ and $S$ followed by a softmax over all the words in the paragraph. The same formula is applied to compute if the word $i$ is the end token by replacing the start vector $S$ with the end vector $E$. Finally, the score of a candidate span from position $i$ to position $j$ is defined as $S \cdot T_{i} + E \cdot T_{j}$, and the maximum scoring span where $j \geq i$ is used as a prediction.


\subsection{Database approach}

The objective of proposing this baseline is to showcase which is the performance of an ad-hoc method using heuristics and commercial software to achieve the best possible performance. Since obtaining a human performance analysis is near impossible
because it would mean that the people involved in the experiment should check more than $14k$ documents for each question, we see this baseline as a performance to beat in a medium-long term.

This approach also breakdown the task in the same retrieval and answering stages. However, in this case the ranking of the results is binary rather indicating if a document is relevant or not. For that, we first run a commercial OCR over the document collection, extracting not only the recognized text, but also the key-value relationship between the field names and their values, including checkboxes. This is followed by a process to correct possible OCR recognition errors for the fields with low variability (field names, parties and reporting options) and normalize all the dates by parsing them to the same format. Finally, we map the key-value pairs into a database like data structure. At the time of answering a question, the fields in the query are compared with those in the stored records. If all the constraints are met, that document is considered relevant and is given a confidence of $1$, while otherwise it is given a confidence of $0$. Finally, the requested value in the question is extracted from the records of the relevant documents. 

It is very important to consider two relevant aspects on this baseline. 
First, it is a very rigid method that does not allow any modification in the data and therefore, is not generalizable at all. Moreover, it requires a preprocessing that is currently done manually to parse the query from Natural Language to a Structured Query Language (SQL). 





\section{Results}

\newcommand\retTsGoogle{71.62}
\newcommand\retTsTextract{72.84}
\newcommand\retDatabase{71.06}

\subsection{Evidences}

To initially assess the retrieval performance of the methods, we first compare two different commercial OCR systems that we are going to use for text spotting, Google OCR~\cite{google2020ocr} and Amazon Textract~\cite{amazon2021textract}. As reported in table \ref{tab:results_retrieval} the performance on text spotting with the latter OCR is better than Google OCR, and is the only one capable of extracting the key-value relations for the database approach. For this reason we use this as the standard OCR for the rest of the text spotting baselines. 

\begin{table}
\centering
\begin{tabular}{lc}
\toprule
\textbf{Retrieval method}  & \textbf{MAP}       \\
\midrule
Text spotting (google)     & \retTsGoogle{}     \\
Text spotting (textract)   & \retTsTextract{}   \\
Database                   & \retDatabase{}     \\
\bottomrule                                     \\
\end{tabular}
\caption{Performance of different retrieval methods.}
\label{tab:results_retrieval}
\end{table}

\begin{figure}
    \centering
    \includegraphics[width=\textwidth]{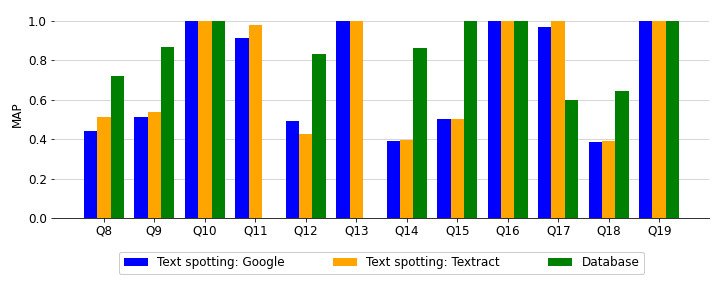}
    \caption{Evidence retrieval performance of the different methods reported by each question in the test set.}
    \label{fig:retrieval_results}
\end{figure}

Compared to text spotting, the database retrieval average performance is similar. However, as depicted in figure~\ref{fig:retrieval_results} we can appreciate that performs better for all the questions but the number $11$ where it gets a MAP of $0$. This is the result from the fact that the key-value pair extractor is not able to capture the relation between some of the forms fields, in this case the treasurer name, and consequently it catastrophically fails at retrieving documents with specific values on those fields, one of the main drawbacks of such rigid methods. On the other hand, the questions where the database approach shows a greater performance gap are those where in order to find the relevant documents the methods must search not only documents with a particular value, but understand more complex constraints such as the ones described in section~\ref{QuestionsAndAnswers}, which are finding documents between two dates (question 9), after a date (question 18), documents that do not contain a particular value (question 12), or where several values are considered as correct (question 14). 

\subsection{Answers}

For the BERT QA method we use the pretrained weights bert-large-uncased-whole-wordmasking-finetuned-squad from the Transformers library~\cite{wolf2020transformers}. This is a pretrained model finetuned on SQuAD 1.1 question answering task consisting on more than $100,000$ questions over $23,215$ paragraphs. Then, we finetune it again on the DocVQA dataset for 2 epochs following~\cite{mathew2021docvqa} to teach the model reason about document concepts as well as adapting the new context style format. Finally, we perform a third finetunning phase on the DocCVQA sample set for $6$ epochs. Notice that the sample set is specially small and during these $6$ epochs the model only see around $80$ samples. Nonetheless, this is sufficient to improve the answering performance without harming the previous knowledge.

Given the collection nature of DocCVQA, the answer to the question usually consists on a list of texts found in different documents considered as relevants. In our case, we consider a document as relevant when the confidence provided for the retrieval method on that document is greater than a threshold. For the text spotting methods we have fixed the threshold through an empirical study 
where we have found that the best threshold is 0.9. In the case of the database approach, given that the confidence provided is either 0 or 1, we consider relevant all positive documents. \\


In the experiments we use the BERT answering baseline to answer the questions over the ranked documents from the text spotting and the database retrieval methods. But we only use the database method to answer the ranked documents from the same retrieval approach. As reported in table~\ref{tab:results} the latter is the one that performs the best. The main reason for this is that the wrong retrieval of the documents prevents the answering methods to find the necessary information to provide the correct answers. Nevertheless, the fact of having the key-value relations allows the database method to directly output the value for the requested field as an answer while BERT needs to learn to extract it from a context that has partially lost the spatial information of the recognized text when at the time of being created, the value of a field might not be close to the field name, losing the semantic connection between the key-value pair. To showcase the answering performance upper bounds of the answering methods we also provide their performance regardless of the retrieval system, where the documents are ranked according to the test ground truth.

\setlength{\tabcolsep}{8pt}  

\begin{table}
\centering
\begin{tabular}{llcc}
\toprule
\textbf{Retrieval}         & \textbf{Answering}         & \multirow{2}{*}{\textbf{MAP}} & \multirow{2}{*}{\textbf{ANLSL}}       \\
\textbf{method}            & \textbf{method}            &                               &                                       \\
\midrule
Text spotting              & BERT                       & \retTsTextract{}  & 0.4513        \\
Database                   & BERT                       & \retDatabase{}    & 0.5411        \\
Database                   & Database                   & \retDatabase{}    & 0.7068        \\
GT                         & BERT                       & 100.00            & 0.5818        \\
GT                         & Database                   & 100.00            & 0.8473        \\
\bottomrule                                                                                 \\
\end{tabular}
\caption{Baselines results comparison.}
\label{tab:results}
\end{table}
\setlength{\tabcolsep}{6pt}  

\begin{figure}
    \centering
    \includegraphics[width=\textwidth]{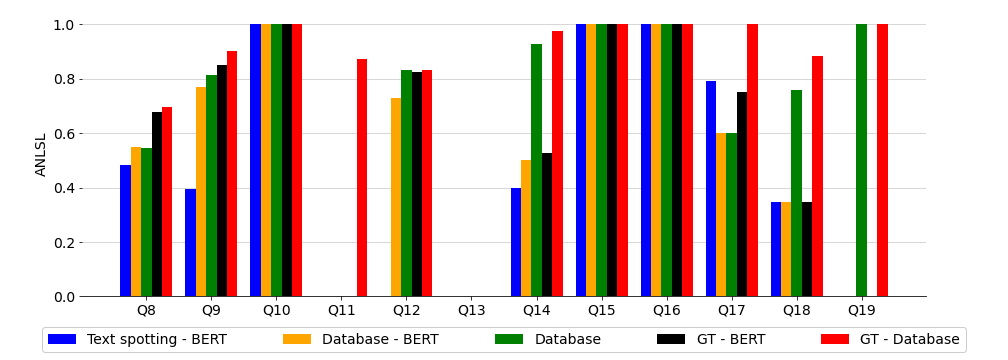}
    \caption{Answering performance of the different methods reported by each question in the test set.} 
    \label{fig:answering_results}
\end{figure}

As depicted in figure~\ref{fig:answering_results}, BERT does not perform well when the answer are candidate's names (questions 8, 9, 11, 14 and 18). However, it has a better performance when asking about dates (questions 16 and 17) or legislative counties (question 10). On the other hand, the database approach is able to provide the required answer, usually depending solely on whether the text and the key-value relationships have been correctly recognized. 

The most interesting question is the number 13, where none of the methods are able to answer the question regardless of a correct retrieval. This question asks if a candidate selected a specific checkbox value. The difference here is that the answer is \emph{No}, in contrast to the sample question number 3. Then, BERT can't answer because it lacks of a document collection point of view, and moreover, since it is an extractive QA method, it would require to have a \emph{No} in the document surrounded with some context that could help to identify that word as an answer. On the other hand, the database method fails because of its logical structure. If there is a relevant document for that question, it will find the field for which the query is asking for, or will answer '\emph{Yes}' if the question is a Yes/No type.


\section{Conclusions and Future Work}
This work introduces a new and challenging task to both the VQA and DAR research fields. We presented the DocCVQA that aims to provide a new perspective to Document understanding and highlight the importance and difficulty of contemplating a whole collection of documents. We have shown the performance of two different approaches. On one hand, a text spotting with an extractive QA baseline that, although it has lower generic performance it is more generic and could achieve similar performance in other types of documents. And on the other hand, a baseline that represents the documents by their key-value relations that despite achieving quite good performance, is still far from being perfect and because of its design is very limited and can't generalize at all when processing other types of documents. In this regard, we believe that the next steps are to propose a method that can reason about the whole collection in a single stage, being able to provide the answer and the positive evidences.

\section*{Acknowledgements}

This work has been supported by the UAB PIF scholarship B18P0070 and the Consolidated Research Group 2017-SGR-1783 from the Research and University Department of the Catalan Government.


\bibliographystyle{splncs04}
\footnotesize
\bibliography{main}

\end{document}